\documentclass[twocolumn,showpacs,preprintnumbers,amsmath,amssymb,floatfix,prb]{revtex4}

\usepackage{graphicx}% Include figure files
\usepackage{dcolumn}% Align table columns on decimal point
\usepackage{bm}% bold math
\usepackage{subfigure}
\usepackage{multirow}

\def\slantfrac#1#2{\hbox{$\,^#1\!/_#2$}}

\def\onethird{\slantfrac{1}{3}}

\begin{document}

\title{First-principles calculation of magnetoelastic coefficients and
  magnetostriction \\in the spinel ferrites CoFe$_{2}$O$_{4}$ and
  NiFe$_{2}$O$_{4}$}

\author{Daniel Fritsch} 
\email{daniel.fritsch@bristol.ac.uk} 
\affiliation{H. H. Wills Physics Laboratory, University of Bristol, Tyndall Avenue, Bristol BS8 1TL, United Kingdom}
\altaffiliation[Previous address: ]{School of Physics, Trinity College, Dublin 2, Ireland}
\author{Claude Ederer} 
\email{claude.ederer@mat.ethz.ch}
\affiliation{Materials Theory, ETH Z\"urich, Wolfgang-Pauli-Strasse
  27, 8093 Z\"urich, Switzerland}
\altaffiliation[Previous address: ]{School of Physics, Trinity College, Dublin 2, Ireland}

\date{\today}

\begin{abstract}
We present calculations of magnetostriction constants for the spinel
ferrites CoFe$_{2}$O$_{4}$ and NiFe$_{2}$O$_{4}$ using density
functional theory within the GGA+$U$ approach. Special emphasis is
devoted to the influence of different possible cation distributions on
the $B$ site sublattice of the inverse spinel structure on the
calculated elastic and magnetoelastic constants. We show that the
resulting symmetry-lowering has only a negligible effect on the
elastic constants of both systems as well as on the magnetoelastic
response of NiFe$_2$O$_4$, whereas the magnetoelastic response of
CoFe$_2$O$_4$ depends more strongly on the specific cation
arrangement. In all cases our calculated magnetostriction constants
are in good agreement with available experimental data. Our work thus
paves the way for more detailed first-principles studies regarding the
effect of stoichiometry and cation inversion on the magnetostrictive
properties of spinel ferrites.
\end{abstract}



\maketitle

\section{Introduction}
\label{Introduction}

Magnetostriction describes the deformation of a ferro- or
ferrimagnetic material during a magnetization
process.\cite{Kittel_RevModPhys21_541,
  Lee_RepProgPhys18_184,Callen_PhysRev129_578,Callen_PhysRev139_A455,Clark1980531,Cullen_MatSciTechnol3B_529,Tremolet_MagnetoelasticEffects}
Thereby, one can distinguish between the \emph{spontaneous volume
  magnetostriction}, which is independent of the magnetic field
direction, and the so-called \emph{linear magnetostriction} which
characterizes the change of length along a certain direction that
depends on the orientation of the applied magnetic field.  The same
magnetoelastic interaction that causes magnetostriction also leads to
changes in the magnetic anisotropy as function of an externally
applied strain.

Magnetostrictive materials are very important for applications as
magnetic field sensors and magneto-mechanical actuators, where a large
(and often also preferably linear) magnetic field response is
essential.\cite{Tremolet_MagnetostrictiveMaterials} On the other hand
magnetostriction also causes noise and frictional losses in magnetic
transformer cores, so that in this context a minimization of
magnetostriction is desirable.

CoFe$_2$O$_4$ (CFO) is known to have one of the largest
magnetostriction among magnetic materials that do not contain any
resource-critical rare-earth elements.\cite{Brabers1995189} It has
thus recently come into focus for use in
magnetostrictive-piezoelectric
composites,\cite{Zheng_Science303_661,Zavaliche_et_al:2005,Dix_et_al:2010}
where the goal is to achieve cross coupling between magnetic and
dielectric degrees of freedom. Due to its insulating character and
high magnetic ordering temperature, CFO together with NiFe$_2$O$_4$
(NFO) and other spinel ferrites is also a very attractive candidate
for spintronics applications, in particular for spin-filtering tunnel
barriers.\cite{Chapline/Wang:2006,Lueders_et_al_APL:2006} For many of
these applications, thin films of CFO and NFO are epitaxially grown on
substrates with different lattice constants. The resulting
substrate-induced strain can then lead to distinctly different
properties of the thin films compared to the corresponding bulk
materials.

In view of this, a good quantitative understanding of magnetoelastic
properties of spinel ferrites, that provides a solid basis for the
interpretation of experimental results and allows for further
optimization of magnetostrictive properties, is highly desirable. In
particular, the ability to accurately predict effects of cation
off-stoichiometry or surface and interface effects can provide
valuable insights into the fundamental mechanisms determining the
observed properties.

In previous work we have shown that first-principles calculations
based on density-functional theory (DFT) provide a suitable
description of the magnetoelastic properties of spinel
ferrites,\cite{Fritsch_PRB82_104117,Fritsch_JPhysConfSer292_012104}
thus demonstrating the feasibility of more detailed studies into
strain-induced effects in thin film structures composed of CFO and
NFO. Here we extend our previous study, in order to provide a more
comprehensive picture of the magnetoelastic response of CFO and NFO,
in particular including first-principles calculations of the complete
set of cubic magnetoelastic and magnetostrictive coefficients. Most
importantly, we investigate the influence of different possible cation
distributions on the spinel $B$ site sublattice on the magnetoelastic
response of these materials. The purpose of the present work is to
provide a first-principles based description of magnetoelastic
coupling in spinel ferrites that can be used as basis for further
studies of the effect of cation substitution or off-stoichiometry on
the magnetostrictive properties of this important class of materials.

This paper is organized as follows. In Sec.~\ref{Chapter2.1} the
spinel crystal structure is discussed, with special emphasis on cation
inversion and different possible cation arrangements on the $B$ site
sublattice. A general overview of magnetoelastic theory in cubic and
tetragonal crystals is given in
Sec.~\ref{Chapter2.2}. Sec.~\ref{Chapter2.3} describes how we
determine all elastic and magneto-elastic coefficients from total
energy electronic structure calculations, while Sec.~\ref{Chapter2.4}
provides some more technical details of our calculations. Our results
for CFO and NFO are presented in Sec.~\ref{Chapter3}, and our main
conclusions are summarized in Sec.~\ref{Chapter4}.

\section{Theoretical background and Computational Method}
\label{Chapter2}

\subsection{Inverse spinel structure and different cation distributions}
\label{Chapter2.1}

\begin{figure}
\includegraphics*[width=0.8\columnwidth]{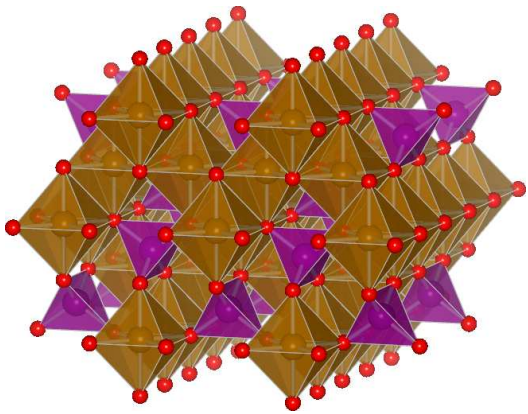}
\caption{(Color online) The spinel structure consists of an fcc
  network of oxygen anions (red) with cations occupying different
  interstitial sites of the fcc lattice, resulting in tetrahedrally
  coordinated $A$ sites (purple) and octahedrally coordinated $B$
  sites (brown). Picture has been generated using VESTA.\cite{VESTA}}
\label{fig:spinel}
\end{figure}

Both CFO and NFO crystallize in the cubic spinel structure (see
Fig.~\ref{fig:spinel}), which belongs to space group $Fd\bar{3}m$
(No. 227). The spinel structure contains two inequivalent cation
sites, a tetrahedrally coordinated $A$ site and an octahedrally
coordinated $B$ site. In the \textit{normal} spinel structure each of
these sites is occupied by a particular cation species (e.g. divalent
Mn$^{2+}$ on the $A$ site and trivalent Fe$^{3+}$ on the $B$ site in
the case of MnFe$_2$O$_4$). However, in the \textit{inverse} spinel
structure, the more abundant cation species (here: Fe$^{3+}$) occupies
all $A$ sites and 50\,\% of the $B$ sites, with the remaining 50\,\%
of $B$ sites occupied by the less abundant cation species (here:
Co$^{2+}$ or Ni$^{2+}$). In practice, intermediate cases can also
occur, characterized by an inversion parameter $\lambda$, ranging from
$\lambda=0$ for the \textit{normal} spinel structure to $\lambda=1$
for complete inversion.

Both CFO and NFO are experimentally found to be inverse spinels, with
$\lambda \approx 1$ for NFO but only incomplete inversion for CFO
(with $\lambda$ between $0.76-0.93$, depending strongly on sample
preparation conditions).\cite{Brabers1995189,Moyer_PRB83_035121} Both
materials are generally found to be perfectly cubic, with a random
distribution of divalent and trivalent cations over the $B$ site
sublattice. However, indications for short-range cation order on the
$B$ sites have been reported recently for the case of NFO, both in
bulk single crystals as well as in thin
films.\cite{Ivanov_PRB82_024104,Iliev_PRB83_014108}

\begin{figure*}
\includegraphics[width=\textwidth,clip]{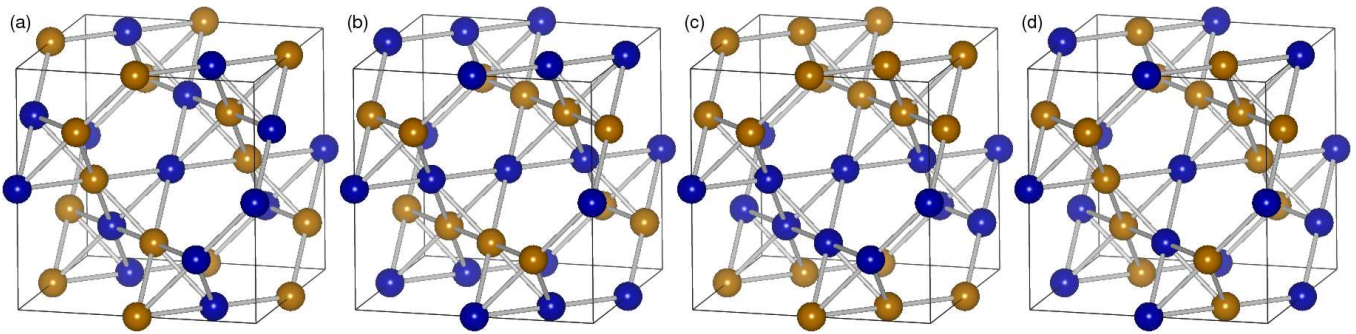}
\caption{\label{FigSpinelCationDistribution}(Color online) Cation
  distribution of Fe (brown) and Co (Ni) (blue) on the $B$ sites of
  the spinel structure for the different configurations used in our
  calculations. Note that only the $B$ sublattice is shown. From left
  to right the depicted structures correspond to spacegroups (a)
  $P4_{1}22$ (No. 91), (b) $Imma$ (No. 74), and (c) $P\bar{4}m2$
  (No. 115). Figure (d) on the right displays the CFO low-energy
  solution with incomplete degree of inversion, $\lambda=0.75$
  corresponding to spacegroup $P1$ (No. 1).\cite{Fritsch_APL99_081916}
  Pictures have been generated using VESTA.\cite{VESTA}}
\end{figure*}

In the present work we represent the inverse spinel structure within a
tetragonal unit cell containing 4 formula units (see also
Ref.~\onlinecite{Fritsch_APL99_081916}) using lattice vectors
$\vec{a}_1 = \left( a/2, -a/2, 0 \right)$, $\vec{a}_2 = \left( a/2,
a/2, 0 \right)$, and $\vec{a}_3 = \left( 0, 0, c \right)$, so that
$c/a=1$ corresponds to the unstrained, nominally cubic case. By
distributing equal amounts of Co (respectively Ni) and Fe on the 8 $B$
sites within this unit cell, 70 cation arrangements belonging to 8
different spacegroups can be generated. In the following we consider
only the three high-symmetry arrangements shown in
Fig.~\ref{FigSpinelCationDistribution}~(a)-(c), plus one additional
low-energy configuration for CFO, corresponding to 75\,\% inversion,
shown in Fig.~\ref{FigSpinelCationDistribution}~(d). The specific
cation arrangements shown in Fig.~\ref{FigSpinelCationDistribution} in
combination with the periodic boundary conditions corresponding to the
tetragonal lattice vectors reduce the space group symmetries to
$P4_{1}22$ (No. 91), $Imma$ (No. 74), and $P\bar{4}m2$ (No. 115) for
the fully inverse configurations, and to $P1$ (No. 1) for the case
with 75~\% inversion. As we have previously
shown,\cite{Fritsch_APL99_081916} both $P4_{1}22$ and $Imma$
correspond to low energy configurations for the fully inverse case,
with $P4_{1}22$ slightly lower in energy than $Imma$ for both CFO and
NFO, whereas the $P\bar{4}m2$ configuration is energetically much less
favorable. The $P1$ structure represents a low energy configuration
for the case $\lambda=0.75$.\cite{Fritsch_APL99_081916} We also note
that the $P4_{1}22$ configuration corresponds to the local structure
suggested for the experimentally observed short-range order in
NFO,\cite{Ivanov_PRB82_024104,Iliev_PRB83_014108} whereas the $Imma$
configuration is equivalent to the one used in our previous study of
magneto-elastic effects in CFO and
NFO.\cite{Fritsch_PRB82_104117,Fritsch_JPhysConfSer292_012104}

\subsection{Magnetoelastic theory}
\label{Chapter2.2}

Within the phenomenological theory of magnetoelasticity, the
magnetoelastic energy density $f=E/V$ is expressed in terms of the
direction cosines of the magnetization vector, $\alpha_i$
($i={x,y,z}$), and the components of the strain tensor
$\varepsilon_{ij}$, relative to a suitably chosen (nonmagnetic)
reference
state.\cite{Kittel_RevModPhys21_541,Lee_RepProgPhys18_184,Callen_PhysRev129_578,Callen_PhysRev139_A455,Clark1980531,Cullen_MatSciTechnol3B_529,Tremolet_MagnetoelasticEffects}
This energy density can be divided into a purely elastic term,
$f_\text{el}$, and a magnetoelastic coupling term, $f_\text{me}$,
which is usually taken as linear in the strain components. For a cubic
crystal these terms have the following form:\cite{footnote1}
\begin{equation}
\label{EqFElCub}
\begin{split}
f_\text{el}^\text{cubic}=&\frac{1}{2}C_{11}(\varepsilon_{xx}^{2}+\varepsilon_{yy}^{2}+\varepsilon_{zz}^{2})
+2C_{44}(\varepsilon_{xy}^{2}+\varepsilon_{yz}^{2}+\varepsilon_{zx}^{2})
\\ +&C_{12}(\varepsilon_{yy}\varepsilon_{zz}+\varepsilon_{xx}\varepsilon_{zz}+\varepsilon_{xx}\varepsilon_{yy})
\,,
\end{split}
\end{equation}
and
\begin{equation}
\label{EqFMeCub}
\begin{split}
f_\text{me}^\text{cubic}=&B_{0}(\varepsilon_{xx}+\varepsilon_{yy}+\varepsilon_{zz})
\\ +&B_{1}(\alpha_{x}^{2}\varepsilon_{xx}+\alpha_{y}^{2}\varepsilon_{yy}+\alpha_{z}^{2}\varepsilon_{zz})
\\ +&2B_{2}(\alpha_{x}\alpha_{y}\varepsilon_{xy}+\alpha_{y}\alpha_{z}\varepsilon_{yz}+\alpha_{z}\alpha_{x}\varepsilon_{zx})
\,,
\end{split}
\end{equation}
where $C_{11}$, $C_{12}$, and $C_{44}$ are elastic and $B_{0}$,
$B_{1}$, and $B_{2}$ are magnetoelastic coupling constants.

The relative length change along an arbitrary (measuring) direction
with direction cosines $\beta_i$ is given by:
\begin{equation}
\frac{\Delta l}{l} = \sum_{i,j} \varepsilon_{ij} \beta_i \beta_j \,,
\end{equation}
where the strain components depend on the magnetization directions.
These equilibrium strains as function of the magnetization direction
can be found by minimizing the sum of the two energy expressions
\eqref{EqFElCub} and \eqref{EqFMeCub} with respect to all strain
components. This results in:
\begin{equation}
\label{EqDeltaLCub}
\begin{split}
\left.\frac{\Delta l}{l}\right|_\text{cubic}=
&\lambda^{\alpha}+\frac{3}{2}\lambda_{100}\left(\alpha_{x}^{2}\beta_{x}^{2}+\alpha_{y}^{2}\beta_{y}^{2}+\alpha_{z}^{2}\beta_{z}^{2}-\onethird\right)\\
+&3\lambda_{111}\left(\alpha_{x}\alpha_{y}\beta_{x}\beta_{y}+\alpha_{y}\alpha_{z}\beta_{y}\beta_{z}+\alpha_{x}\alpha_{z}\beta_{x}\beta_{z}\right)\,.
\end{split}
\end{equation}
Here, $\lambda^{\alpha}=-(B_{0}+B_{1}/3)/(C_{11}+2C_{12})$ describes a
pure volume magnetostriction that is independent of the magnetization
direction (this term is sometimes omitted from the above formula and
is of no concern in the present work). The widely used
magnetostriction constants of a cubic crystal are given by:
\begin{equation}
\label{EqLambda100Cub}
\lambda_{100}=-\frac{2}{3}\frac{B_{1}}{C_{11}-C_{12}} \,,
\end{equation}
and
\begin{equation}
\label{EqLambda111Cub}
\lambda_{111}=-\frac{B_{2}}{3C_{44}} \,.
\end{equation}
These two coefficients measure the fractional length change along the
[100] ($\beta_{x}=1, \beta_{y}=\beta_{z}=0$) and [111]
($\beta_{i}=1/\sqrt{3}$) directions, when the sample is magnetized to
saturation along the [100] ($\alpha_{x}=1, \alpha_{y}=\alpha_{z}=0$)
and [111] ($\alpha_{i}=1/\sqrt{3}$) directions, relative to an ideal
demagnetized reference state which is defined by
$\left<\alpha_{i}^{2}\right>=\onethird$ and
$\left<\alpha_{i}\alpha_{j}\right>=0$. In a polycrystalline sample one
can only measure a direction average over both $\lambda_{100}$ and
$\lambda_{111}$ given by:\cite{Lee_RepProgPhys18_184}
\begin{equation}
\label{EqLambdaS}
\lambda_{S}=\frac{2}{5}\lambda_{100}+\frac{3}{5}\lambda_{111} \,.
\end{equation}

As noticed in Sec.~\ref{Chapter2.1}, the cation arrangements used to
describe the inverse spinel structure within our calculations lower
the cubic symmetry of the ideal spinel structure to tetragonal
($P4_122$ and $P\bar{4}m2$), orthorhombic ($Imma$), or even triclinic
($P1$). A full first-principles description of magnetoelastic effects
within these lower symmetries would require the calculation of 6 (9,
21) different elastic and 7 (12, 36) magnetoelastic coupling constants
for the mentioned tetragonal (orthorhombic, triclinic) spacegroups,
respectively.\cite{Callen_PhysRev139_A455} Due to the resulting large
computational effort, and considering the fact that experimentally
both CFO and NFO are found to be cubic, we do not attempt such a full
determination of all elastic and magnetoelastic coefficients within
the lower symmetries, and instead evaluate our results using the
relations for the cubic case described above (i.e., similar to our
previous work in Refs.~\onlinecite{Fritsch_PRB82_104117} and
\onlinecite{Fritsch_JPhysConfSer292_012104}). To estimate the degree to
which the lower symmetry affects our calculated coefficients, we also
compare some of our data to the correct formulas corresponding to the
lower symmetry. For simplicity we hereby restrict ourselves to the
tetragonal case. The required equations are presented in the
following.

Within the lower tetragonal symmetry there are six independent elastic
and seven different magnetoelastic coupling constants, in contrast to
the three elastic and three magnetoelastic coefficients in the cubic
case.\cite{Callen_PhysRev139_A455} The resulting expressions for
$f_\text{el}$ and $f_\text{me}$ then
read:\cite{Cullen_MatSciTechnol3B_529}
\begin{equation}
\label{EqFElTet}
\begin{split}
f_{\text{el}}^{\text{tet}} &=
\frac{1}{2}c_{11}\left(\varepsilon_{xx}^{2}+\varepsilon_{yy}^{2}\right)
+\frac{1}{2}c_{33}\varepsilon_{zz}^{2}\\ &+
c_{12}\varepsilon_{xx}\varepsilon_{yy}+c_{13}\left(\varepsilon_{xx}+\varepsilon_{yy}\right)\varepsilon_{zz}
\\ &+
2c_{44}\left(\varepsilon_{yz}^{2}+\varepsilon_{xz}^{2}\right)+2c_{66}\varepsilon_{xy}^{2}\,,
\end{split}
\end{equation}
with $c_{ij}$ denoting the six different tetragonal elastic constants,
and
\begin{equation}
\label{EqFMeTet}
\begin{split}
f_{\text{me}}^{\text{tet}}=&b_{11}\left(\varepsilon_{xx}+\varepsilon_{yy}\right)+b_{12}\varepsilon_{zz}
\\ +&b_{21}\left(\alpha_{z}^{2}-\onethird\right)\left(\varepsilon_{xx}+\varepsilon_{yy}\right)+b_{22}\left(\alpha_{z}^{2}-\onethird\right)\varepsilon_{zz}
\\ +&\frac{1}{2}b_{3}\left(\alpha_{x}^{2}-\alpha_{y}^{2}\right)\left(\varepsilon_{xx}-\varepsilon_{yy}\right)+b_{3}^{'}\alpha_{x}\alpha_{y}\varepsilon_{xy}
\\ +&b_{4}\left(\alpha_{x}\alpha_{z}\varepsilon_{xz}+\alpha_{y}\alpha_{z}\varepsilon_{yz}\right)
\,,
\end{split}
\end{equation}
with the various $b$'s denoting the seven different tetragonal
magnetoelastic coupling constants. The corresponding cubic expressions
\eqref{EqFElCub} and \eqref{EqFMeCub} can then be obtained from
\eqref{EqFElTet} and \eqref{EqFMeTet} with the additional symmetry
constraints: $c_{11}=c_{33}=C_{11}$, $c_{12}=c_{13}=C_{12}$,
$c_{44}=c_{66}=C_{44}$, $b_{11}=b_{12}=B_{0}+\onethird B_{1}$,
$b_{22}=-2b_{21}=b_{3}=B_{1}$, and $b_{3}^{'}=b_{4}=B_{2}$.

\subsection{Determination of elastic and magnetoelastic constants}
\label{Chapter2.3}

In order to determine the (cubic) elastic constants for CFO and NFO,
we first perform a full structural relaxation of both systems. Similar
to our previous
investigations,\cite{Fritsch_PRB82_104117,Fritsch_JPhysConfSer292_012104,Fritsch_APL99_081916}
we thereby constrain the lattice vectors to ``cubic'' symmetry
($c/a=1$) and fix the internal coordinates of the $A$ and $B$ cations
to ideal values corresponding to the cubic spinel structure, i.e., we
only allow for an optimization of the total volume and the oxygen
positions. We then determine the three independent cubic elastic
constants $C_{11}$, $C_{12}$, and $C_{44}$, and the two cubic
magnetoelastic coupling constants $B_1$ and $B_2$ by distorting the
equilibrium crystal structure in three different ways: i) isotropic
volume expansion, ii) constraining two of the three lattice dimensions
and relaxing the third (``epitaxial strain''), and iii) by applying a
volume-conserving shear strain.

\noindent i) \textit{Isotropic volume expansion.} The dependence of
the total energy $E_{\text{tot}}$ on the unit cell volume $V$ provides
the bulk modulus $B$, which is defined as
\begin{equation}
\label{EqBulkModulus}
B=V_{0}\left.\left(\frac{\partial^{2}E_{\text{tot}}}{\partial V^{2}}\right)\right|_{(V=V_{0})}\,,
\end{equation}
with $V_{0}$ being the equilibrium volume. According to
Eq.~\eqref{EqFElCub} the bulk modulus $B$ of a cubic crystal can be
expressed in terms of the elastic moduli $C_{11}$ and $C_{12}$ as
follows:
\begin{equation}
\label{EqBulkModulusElasticModuli}
B=\frac{1}{3}\left(C_{11}+2C_{12}\right) \,.
\end{equation}

\noindent ii) \textit{Epitaxial strain.} We follow the approach of
Ref.~\onlinecite{Fritsch_PRB82_104117} to obtain a second independent
elastic constant by applying \emph{epitaxial strain}, i.e., we
constrain the ``in-plane'' lattice constant to values ranging from
$-4\%$ to $+4\%$ relative to the theoretical equilibrium lattice
constant $a_0$, and we relax the ``out-of-plane'' lattice constant and
all internal structural parameters of the oxygen anions. The relation
between the relaxed out-of-plane strain $\varepsilon_{\perp}$ and the
fixed in-plane strain $\varepsilon_{||}$ then defines the so-called
two-dimensional Poisson ratio $\nu_{2D}$. It follows from
Eq.~(\ref{EqFElCub}) that for a cubic system $\nu_{2D}$ is given as:
\begin{equation}
\label{EqPoissonRatio}
\nu_{2D}=-\frac{\varepsilon_{\perp}}{\varepsilon_{||}}=2\frac{C_{12}}{C_{11}}\,.
\end{equation}

The elastic moduli $C_{11}$ and $C_{12}$ can then be obtained from
Eqs.~\eqref{EqBulkModulusElasticModuli} and \eqref{EqPoissonRatio}
using the bulk modulus and two-dimensional Poisson ratio calculated
from DFT.

For the cation arrangements with tetragonal, orthorhombic, or
triclinic symmetry depicted in Fig.~\ref{FigSpinelCationDistribution}
the ratio $\varepsilon_{\perp}/\varepsilon_{||}$ can be different for
different orientations of ``out-of-plane'' and ``in-plane'' directions
relative to the crystal axes. To quantify the resulting difference we
perform calculations for two symmetry-inequivalent orientations of the
applied strain $\varepsilon_{||}$. In particular we apply the
epitaxial constraint first within the $xy$ plane
($\varepsilon_{||}=\varepsilon_{xx}=\varepsilon_{yy}$ and
$\varepsilon_{\perp}=\varepsilon_{zz}$) and then also within the $yz$
plane ($\varepsilon_{||}=\varepsilon_{yy}=\varepsilon_{zz}$ and
$\varepsilon_{\perp}=\varepsilon_{xx}$). Using the tetragonal energy
expressions of Eqs.~\eqref{EqFElTet} and \eqref{EqFMeTet} together
with the definition of $\nu_{2D}$ in Eq.~\eqref{EqPoissonRatio} one
obtains $\nu_{2D}^{(xy)}=2c_{13}/c_{33}$ and
$\nu_{2D}^{(yz)}=(c_{12}+c_{13})/c_{11}$ for these two cases.
The difference between these two values for $\nu_{2D}$ thus gives a
measure for the difference between $c_{11}$ and $c_{33}$ as well as
between $c_{12}$ and $c_{13}$.

To obtain the magnetoelastic coupling coefficient $B_1$ we monitor the
total energy differences for different orientations of the
magnetization as a function of the applied in-plane constraint
$\varepsilon_{||}$ and relaxed out-of-plane strain
$\varepsilon_\perp=-\nu_{2D} \varepsilon_{||}$. Using the cubic
expression \eqref{EqFMeCub} for $f_\text{me}$ one can see that the
strain dependence of the energy density for all in-plane orientations
of the magnetization is given by $B_1 \cdot \varepsilon_{||}$, whereas
the strain dependence for out-of-plane orientation is given by $-B_1
\cdot \nu_{2D} \cdot \varepsilon_{||}$. The strain dependence of the
total energy difference between out-of-plane versus in-plane
orientation of the magnetization is thus given by:\cite{footnote2}
\begin{equation}
\label{EqDeltaEB1}
\Delta E/V = -(\nu_{2D} + 1 ) B_1 \varepsilon_{\parallel}\,.
\end{equation}
The coefficient $B_1$ can therefore be obtained from the calculated
strain-dependent magnetic anisotropy energies (MAEs) and the
previously determined two-dimensional Poisson ratio $\nu_{2D}$. While
$B_{1}$ is not directly accessible by experimental investigations, it
is related to the magnetostriction constant $\lambda_{100}$ via
Eq.~\eqref{EqLambda100Cub}.

In the tetragonal case the monitored strain dependence of the total
energy difference between out-of-plane versus in-plane directions of
the magnetization will depend on the orientation of ``out-of-plane''
and ``in-plane'' directions with respect to the tetragonal crystal
axes. For the epitaxial constraint applied within the $xy$ plane
(i.e. $\varepsilon_{\parallel}=\varepsilon_{xx}=\varepsilon_{yy}$,
leading to a Poisson ratio $\nu_{2D}^{(xy)}=2c_{13}/c_{33}$) and using
the tetragonal energy density (Eqs.~\eqref{EqFElTet} and
\eqref{EqFMeTet}), the following expression for the strain dependence
of the total energy difference between in-plane and out-of-plane
magnetization can be obtained:
\begin{equation}
\label{EqDeltaETetxy}
(\Delta E)^{(xy)}/V = (2b_{21}-\nu_{2D}^{(xy)}b_{22})\varepsilon_{\parallel}\,,
\end{equation}
which is valid for all in-plane orientations of the magnetization. In
contrast, for the epitaxial constraint applied within the $yz$ plane
(i.e. $\varepsilon_{||}=\varepsilon_{yy}=\varepsilon_{zz}$, leading to
a Poisson ratio $\nu_{2D}^{(yz)}=(c_{12}+c_{13})/c_{11}$) the
resulting $(\Delta E)^{(yz)}/V$ depends on the specific in-plane
direction and is given by:
\begin{widetext}
\begin{equation}
\label{EqDeltaETetyz}
(\Delta E)^{(yz)}/V=
\begin{cases}
\left(-\frac{1}{2}b_{3}(\nu_{2D}^{(yz)}+1)-(b_{21}+b_{22}-\nu_{2D}^{(yz)} b_{21})\right)\varepsilon_{\parallel}& \text{for $(\Delta E)^{(yz)}=E_{100}-E_{001}$} \\
\left(-b_{3}(\nu_{2D}^{(yz)}+1)\right)\varepsilon_{\parallel}& \text{for $(\Delta E)^{(yz)}=E_{100}-E_{010}$} \\
\left(-\frac{3}{4}b_{3}(\nu_{2D}^{(yz)}+1)-\frac{1}{2}(b_{21}+b_{22}-\nu_{2D}^{(yz)} b_{21})\right)\varepsilon_{\parallel}& \text{for $(\Delta E)^{(yz)}=E_{100}-E_{011/01\bar{1}}$}\,.
\end{cases}
\end{equation}
\end{widetext}

\noindent iii) \textit{Volume-conserving shear strain.} The third
cubic elastic modulus $C_{44}$ is calculated according to
Mehl,\cite{Mehl_PRB47_2493} by applying a volume-conserving monoclinic
shear strain in the $xy$ plane
($\varepsilon_{\parallel}=\varepsilon_{xy}$,
$\varepsilon_{\perp}=\varepsilon_{zz}=\varepsilon_{\parallel}^{2}/(1-\varepsilon_{\parallel}^{2})$,
$\varepsilon_{xx}=\varepsilon_{yy}=\varepsilon_{yz}=\varepsilon_{zx}=0$).
The resulting change in total energy can then be written as:
\begin{equation}
E(\pm\varepsilon_{\parallel}) = 2 V C_{44}\varepsilon_{\parallel}^{2} + O[\varepsilon_{\parallel}^{4}] \,,
\end{equation}
which allows for a straight-forward determination of $C_{44}$.

For the cation arrangements with tetragonal, orthorhombic, or
triclinic symmetry depicted in Fig.~\ref{FigSpinelCationDistribution}
different shear planes ($\varepsilon_{xy}$, $\varepsilon_{yz}$,
$\varepsilon_{zx}$) are connected to different elastic moduli
$c_{ii}$. Using the tetragonal energy expressions of
Eqs.~\eqref{EqFElTet} and \eqref{EqFMeTet} together with the
volume-conserving monoclinic strain in the $xy$ plane described above,
one notices the connection of $\varepsilon_{xy}$ and
$c_{66}$. However, choosing a volume-conserving monoclinic strain in
the $yz$ plane ($\varepsilon_{\parallel}=\varepsilon_{yz}$,
$\varepsilon_{\perp}=\varepsilon_{xx}=\varepsilon_{\parallel}^{2}/(1-\varepsilon_{\parallel}^{2})$,
$\varepsilon_{yy}=\varepsilon_{zz}=\varepsilon_{xy}=\varepsilon_{zx}=0$)
yields directly $c_{44}$, allowing for a comparison with $c_{66}$.

Similar to the first magnetoelastic coupling constant $B_1$, the
second coefficient $B_{2}$ is determined by monitoring the total
energy differences between different orientations of the magnetization
as a function of the applied strain $\varepsilon_{\parallel}$.
Depending on whether the shear strain $\varepsilon_{\parallel}$ is
applied within the $xy$ or $yz$ plane, we consider the following
energy differences:
\begin{align}
\label{eq:mae-shear1}
(\Delta E)^{(xy)} &= E_{110}-E_{100/010} = E_{100/010}-E_{1\bar{1}0}\\
(\Delta E)^{(yz)} &= E_{011}-E_{010/001} = E_{010/001}-E_{01\bar{1}}\,.
\label{eq:mae-shear2}
\end{align}
In all cases, the strain-dependence of these total energy differences
can be written as:
\begin{equation}
\label{EqDeltaEB2}
\Delta E/V = B_{2}\varepsilon_{\parallel}\,.
\end{equation}
Thus, the strain dependence of these energy differences is governed by
the magnetoelastic coupling constant $B_2$, which can be determined
from the calculated $\Delta E/V(\epsilon_\parallel)$.

Similar to $B_1$, the magnetoelastic coupling constant $B_2$ is also
not directly accessible by experiment, but it is related to the
magnetostriction constant $\lambda_{111}$ via
Eq.~\eqref{EqLambda111Cub}. Once the magnetostriction constants
$\lambda_{100}$ and $\lambda_{111}$ are obtained, the average
magnetostriction constant $\lambda_{S}$, suitable for polycrystalline
samples, can be calculated from Eq.~\eqref{EqLambdaS}.

\subsection{Other computational details}
\label{Chapter2.4}

All calculations presented in this work are performed using the
projector-augmented wave (PAW) method,\cite{Bloechl_PRB50_17953}
implemented in the Vienna \textit{ab initio} simulation package (VASP
4.6).\cite{Kresse_PRB47_558,Kresse_PRB49_14251,Kresse_CompMatSci6_15,Kresse_PRB54_11169}
Standard PAW potentials supplied with VASP were used in the
calculations, contributing nine valence electrons per Co
(4s$^2$3d$^7$), 16 valence electrons per Ni (3p$^6$4s$^2$3d$^8$), 14
valence electrons per Fe (3p$^6$4s$^2$3d$^6$), and 6 valence electrons
per O (2s$^2$2p$^4$).

The generalized gradient approximation according to Perdew, Burke, and
Ernzerhof (PBE)\cite{Perdew/Burke/Ernzerhof:1996} is used in
combination with the Hubbard ``+$U$''
correction,\cite{Anisimov/Aryatesiawan/Liechtenstein:1997} where
$U$=3~eV and $J$=0~eV is applied to the $d$ states on all transition
metal cations. We have shown in
Refs.~\onlinecite{Fritsch_PRB82_104117},
\onlinecite{Fritsch_JPhysConfSer292_012104}, and
\onlinecite{Fritsch_APL99_081916} that this gives a realistic
description of the electronic structure of CFO and NFO and leads to
results which are in good overall agreement with available
experimental data.

All structural relaxations are performed within a scalar-relativistic
approximation, whereas spin-orbit coupling is included for the
calculation of the MAEs. A plane wave energy cutoff of 500~eV is used,
and the Brillouin zone is sampled using a $\Gamma$-centered 5 $\times$
5 $\times$ 3 $k$-point grid both for the structural optimization and
for all total energy calculations. We have verified that all
quantities of interest, in particular the magnetic anisotropy
energies, are well converged for this $k$-point grid and planewave
energy cutoff.

\section{Results and discussion}
\label{Chapter3}

\subsection{Structural properties}
\label{Chapter3.1}

\begin{table*}
\caption{\label{TableElasticProperties} Optimized equilibrium lattice
  constant $a_0$, bulk modulus $B$, two-dimensional Poisson ratio
  $\nu_{2D}$, and elastic moduli $C_{11}$, $C_{12}$, and $C_{44}$ for
  CFO and NFO, obtained for different cation arrangements and strain
  orientations ($\varepsilon_{\perp}=\varepsilon_{zz}=z$ and
  $\varepsilon_{\perp}=\varepsilon_{xx}=x$)
%(\mbox{``$z${''} $\widehat{=}\, \varepsilon_{\perp} = \varepsilon_{zz}$} and \mbox{``$x$'' $\widehat{=}\, \varepsilon_{\perp} = \varepsilon_{xx}$})
  in comparison to experimental data. The experimental $\nu_{2D}$ has
  been evaluated from Eq.~\eqref{EqPoissonRatio} using the
  experimental elastic constants. $P1$ in case of CFO refers to the
  low-energy solution with incomplete degree of inversion,
  $\lambda=0.75$.\cite{Fritsch_APL99_081916}}
\begin{ruledtabular}
\begin{tabular}{l|cc|ccccc}
\multirow{2}{*}{CFO} & $a_0$ & $B$ &
\multirow{2}{*}{$\varepsilon_{\perp}$} & \multirow{2}{*}{$\nu_{2D}$} & $C_{11}$ & $C_{12}$ & $C_{44}$ \\ 
& (\AA{}) & (GPa) & & & (GPa) & (GPa) & (GPa) \\ 
\hline \multirow{2}{*}{$Imma$} & \multirow{2}{*}{8.463} & \multirow{2}{*}{172.3} & z & 1.132 & 242.5 & 137.3 & 94.9 \\ 
& & & x & 1.147 & 240.8 & 138.1 & 83.2 \\ 
\multirow{2}{*}{$P4_{1}22$} & \multirow{2}{*}{8.464} & \multirow{2}{*}{170.8} & z & 1.129 & 240.7 & 135.9 & 84.7 \\ 
& & & x & 1.147 & 238.7 & 136.9 & $-$ \\ 
\multirow{2}{*}{$P\bar{4}m2$} & \multirow{2}{*}{8.473} & \multirow{2}{*}{168.0} & z & 1.132 & 236.4 & 133.8 & 92.3 \\ 
& & & x & 1.128 & 236.8 & 133.6 & $-$ \\ 
\multirow{2}{*}{$P1$} & \multirow{2}{*}{8.477} & \multirow{2}{*}{167.8} & z & 1.155 & 233.6 & 134.9 & 87.7 \\ 
& & & x & 1.146 & 234.6 & 134.4 & $-$ \\ 
\hline 
Exp. (Ref.~\onlinecite{Li_JMaterSci26_2621}) & 8.392 & 185.7 & & 1.167 & 257.1 & 150.0 & 85.3 \\[2mm]

\multirow{2}{*}{NFO} & $a_0$ & $B$ & \multirow{2}{*}{$\varepsilon_{\perp}$} & \multirow{2}{*}{$\nu_{2D}$} & $C_{11}$ & $C_{12}$ & $C_{44}$ \\
& (\AA{}) & (GPa) & & & (GPa) & (GPa) & (GPa) \\
\hline
\multirow{2}{*}{$Imma$} & \multirow{2}{*}{8.426} & \multirow{2}{*}{177.1} & z & 1.106 & 252.2 & 139.5 & 93.2 \\
& & & x & 1.115 & 251.2 & 140.0 & 87.6 \\
\multirow{2}{*}{$P4_{1}22$} & \multirow{2}{*}{8.428} & \multirow{2}{*}{175.4} & z & 1.116 & 248.7 & 138.8 & 87.4 \\
& & & x & 1.116 & 248.7 & 138.8 & $-$ \\
\multirow{2}{*}{$P\bar{4}m2$} & \multirow{2}{*}{8.435} & \multirow{2}{*}{173.3} & z & 1.116 & 245.7 & 137.1 & 91.0 \\
& & & x & 1.102 & 247.3 & 136.3 & $-$ \\
\hline
Exp. (Ref.~\onlinecite{Li_JMaterSci26_2621}) & 8.339 & 198.2 & & 1.177 & 273.1 & 160.7 & 82.3 \\
\end{tabular}
\end{ruledtabular}
%\footnotetext[1]{Calculated for the small unit cell, see text.} 
%\footnotetext[2]{Low-energy solution for inversion parameter $\lambda=0.75$.\cite{Fritsch_APL99_081916}}
\end{table*}

The equilibrium lattice constants $a_{0}$, bulk moduli $B$,
two-dimensional Poisson ratios $\nu_{2D}$, the resulting elastic
constants $C_{11}$ and $C_{12}$, as well as $C_{44}$ obtained for the
different cation arrangements for both CFO and NFO are given in
Tab.~\ref{TableElasticProperties}. One notices that the calculated
lattice constants for the two low-energy configurations $Imma$ and
$P4_122$ are very similar to each other, and that the ones for the
higher energy $P\bar{4}m2$ configuration and for the case with 75\,~\%
inversion for CFO are slightly larger than that (by less than
0.2\,\%). This increase in lattice constant is mirrored by a
corresponding decrease in the bulk modulus (by about 3\,\%). Overall,
the variation of both bulk modulus and equilibrium lattice constant
between different cation distributions is much smaller than the slight
under- and overestimation of these quantities with respect to the
experimental value, which is within the usual limits of the PBE+$U$
approach (see also Ref.~\onlinecite{Fritsch_PRB82_104117}).

It can also be seen that the difference in the two-dimensional Poisson
ratios obtained for two different orientations of $\varepsilon_\perp$
is rather small and of similar magnitude as the differences between
the various cation arrangements. This indicates that the
symmetry-lowering due to the different cation arrangements has only a
small effect on the elastic properties, which can still to a good
approximation be described by cubic elastic constants $C_{11}$ and
$C_{12}$.

Applying the volume-conserving monoclinic strain as described in
Sec.~\ref{Chapter2.2} yields the remaining elastic modulus $C_{44}$
which is in very good agreement with the experimental values for both
CFO and NFO. To evaluate the influence of different orientations of
$\varepsilon_{\perp}$ on $C_{44}$ we applied
$\varepsilon_{\perp}=\varepsilon_{xy}$ and
$\varepsilon_{\perp}=\varepsilon_{yz}$ with the respective
$\varepsilon_{\parallel}$ to the low energy orthorhombic $Imma$
symmetry. The difference in the obtained $C_{44}$ is slightly larger
compared to the difference in the $C_{11}$ and $C_{12}$, but still
within the typical uncertainties of first-principles methods. 

Overall it appears that while the agreement between the calculated and
experimental lattice constants and elastic moduli is quite good and
within the typical uncertainties of state-of-the-art first-principles
methods, the uncertainties resulting from the symmetry-lowering cation
arrangements are significantly smaller than that. Therefore, the
elastic properties of the various cation arrangements of lower
symmetry can be well described by cubic elastic constants.

\subsection{Magnetoelastic properties}
\label{Chapter3.2}

\subsubsection{NFO}
\label{Chapter3.2.1}

\begin{figure}
\includegraphics[width=\columnwidth,clip]{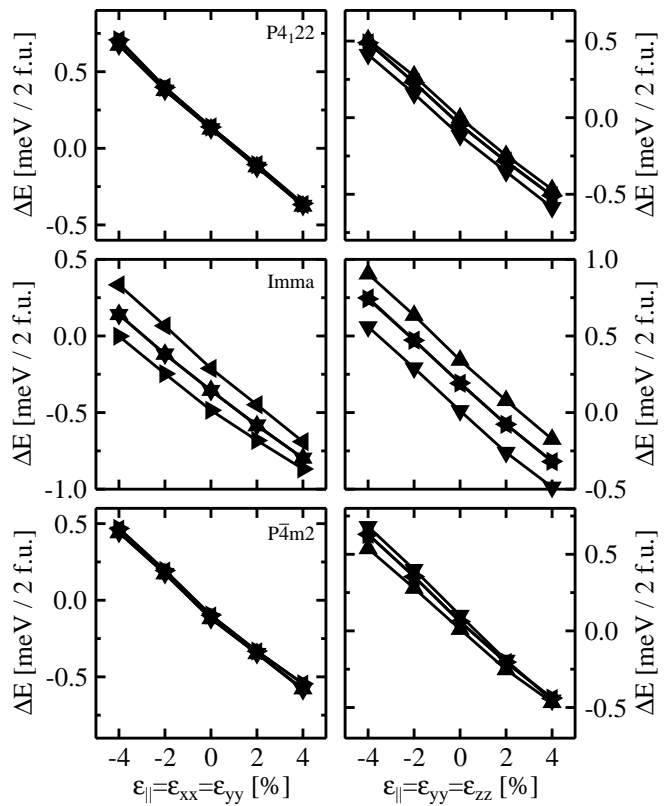}
\caption{\label{FigNFO_MAE_B1} Total energy difference $\Delta E$ per
  two formula units (f.u.) of NFO as function of the epitaxial
  constraint $\varepsilon_{\parallel}$ for different cation
  arrangements. The left (right) panels correspond to the case with
  $\varepsilon_{\perp}=\varepsilon_{zz}$
  ($\varepsilon_{\perp}=\varepsilon_{xx}$). The panels from top to
  bottom refer to symmetries $P4_{1}22$, $Imma$, and $P\bar{4}m2$,
  respectively. In case of $\varepsilon_{\perp}=\varepsilon_{zz}$
  ($\varepsilon_{\perp}=\varepsilon_{xx}$) the depicted energy
  difference $\Delta E$ is taken with respect to the [001] ([100])
  direction, with the symbols denoting $\blacktriangle$ [100] ([010]),
  $\blacktriangledown$ [010] ([001]), $\blacktriangleleft$
  [1$\bar{1}$0] ([01$\bar{1}$]), and $\blacktriangleright$ [110]
  ([011]), respectively.}
\end{figure}

Next we focus on the magnetoelastic coupling in NFO. The calculated
MAEs necessary to determine the magnetoelastic coupling constant
$B_{1}$ are depicted in Fig.~\ref{FigNFO_MAE_B1}. As described in
Sec.~\ref{Chapter2.3} these MAEs are defined here as the energy
differences for various orientations of the magnetization with respect
to the magnetization direction perpendicular to the applied strain
plane, i.e., [001] for
$\varepsilon_{\parallel}=\varepsilon_{xx}=\varepsilon_{yy}$ and [100]
for
$\varepsilon_{\parallel}=\varepsilon_{yy}=\varepsilon_{zz}$. According
to Eq.~\eqref{EqDeltaEB1} the slope of the curves given in
Fig.~\ref{FigNFO_MAE_B1} is directly related to the magnetoelastic
coupling constants $B_{1}$. At first sight, the slopes of all curves
in all panels are very similar and negative, thus leading to a
positive $B_{1}$ (the range of the $y$ axes is the same in all panels
to allow for a direct inspection of slope differences).

\begin{table}
\caption{\label{TableMagneticPropertiesNFO} Magnetoelastic coupling
  constants ($B_1$, $B_2$) and magnetostriction constants
  ($\lambda_{100}$, $\lambda_{111}$, $\lambda_{S}$) for NFO using
  different cation arrangements and strain planes according to
  $\varepsilon_{\perp}=\varepsilon_{zz}=z$
  ($\varepsilon_{\perp}=\varepsilon_{xx}=x$) in comparison with
  available experimental data. The average magnetostriction constant
  $\lambda_{S}$ has been obtained using Eq.~\eqref{EqLambdaS}.}
\begin{ruledtabular}
\begin{tabular}{lcccccccc}
& \multirow{2}{*}{$\varepsilon_{\perp}$} & B$_{1}$ & $\lambda_{100}$ & B$_{2}$ & $\lambda_{111}$ & $\lambda_{S}$ \\ 
& & (MPa) & ($\times 10^{-6}$) & (MPa) & \multicolumn{2}{c}{($\times 10^{-6}$)} \\ 
\hline
\multirow{2}{*}{$P4_{1}22$} & z & 6.6 & $-$40.1 & 2.5 & $-$9.7 & $-$21.9 \\
& x & 6.4 & $-$38.6 & $-$ & $-$ & $-$ \\
\multirow{2}{*}{$Imma$} & z & 6.1 & $-$35.9 & 0.9 & $-$3.4 & $-$16.4 \\
& x & 6.7 & $-$40.3 & 1.9 & $-$7.3 & $-$20.5 \\
\multirow{2}{*}{$P\bar{4}m2$} & z & 6.5 & $-$40.0 & 1.4 & $-$5.3 & $-$19.2 \\
& x & 6.9 & $-$41.3 & $-$ & $-$ & $-$ \\
\hline
\multirow{3}{*}{Exp.} & Ref.~\onlinecite{Bozorth_PhysRev99_1788}\footnotemark[1] & & $-$36.0 & &  $-$4.0 & $-$16.8 \\
         & Ref.~\onlinecite{Smith_JAP37_1001}\footnotemark[2] & & $-$50.9 & & $-$23.8 & $-$34.6 \\
         & Ref.~\onlinecite{Arai_JPhysChemSol36_463}\footnotemark[3] & & $-$43.0 & & $-$20.1 & $-$29.3 \\
\end{tabular}
\end{ruledtabular}
\footnotetext[1]{Single crystals with Ni$_{0.8}$Fe$_{2.2}$O$_4$ composition.}
\footnotetext[2]{Single crystals of NiFe$_2$O$_4$.}
\footnotetext[3]{Single crystals of NiFe$_2$O$_4$.}
\end{table}

In the tetragonal symmetries ($P4_{1}22$ and $P\bar{4}m2$) all curves
fall on top of each other for $\varepsilon_{\perp}=\varepsilon_{zz}$
(left panels), whereas there is a small offset between the curves in
all other cases, due to the lower symmetry. In the even lower $Imma$
symmetry this offset is also present for
$\varepsilon_{\perp}=\varepsilon_{zz}$. Nevertheless, the variation
with strain is very similar in all cases, and the values for $B_{1}$,
obtained by averaging over all curves corresponding to the same
symmetry and strain orientation, are given in
Tab.~\ref{TableMagneticPropertiesNFO}. These values range from 6.1~MPa
to 6.9~MPa, depending on the specific cation arrangement and strain
orientation. Due to these rather small variations, we can conclude
that the magnetostrictive response in NFO can to a good approximation
be described as cubic.

Together with the respective elastic constants from
Tab.~\ref{TableElasticProperties} the magnetostriction constants
$\lambda_{100}$ can be obtained via Eq.~\eqref{EqLambda100Cub}, and
are also listed in Table~\ref{TableMagneticPropertiesNFO}. It can be
seen that there is only a weak influence of either cation arrangement
or different strain planes on the NFO magnetostriction constant
$\lambda_{100}$, which ranges from $-35.9\times 10^{-6}$ to
$-41.3\times 10^{-6}$. This agrees perfectly with experimental data
ranging from $-36.0\times 10^{-6}$ to $-50.9\times 10^{-6}$.

\begin{figure*}
\includegraphics[width=0.8\textwidth,clip]{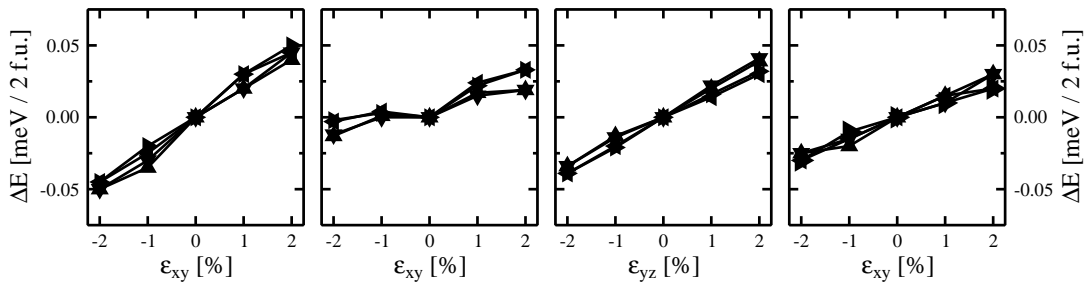}
\caption{\label{FigNFO_MAE_B2} Total energy difference $\Delta E$ per
  two formula units (f.u.) of NFO as function of shear strain for
  different cation arrangements. The panels from left to right refer
  to symmetries $P4_{1}22$
  ($\varepsilon_{\parallel}=\varepsilon_{xy}$), $Imma$
  ($\varepsilon_{\parallel}=\varepsilon_{xy}$), $Imma$
  ($\varepsilon_{\parallel}=\varepsilon_{yz}$), and $P\bar{4}m2$
  ($\varepsilon_{\parallel}=\varepsilon_{xy}$), respectively. The
  depicted energy differences $\Delta E$ correspond to [110]-[100]
  ($\blacktriangle$), [110]-[010] ($\blacktriangledown$),
  [100]-[1$\bar{1}$0] ($\blacktriangleright$), and [010]-[1$\bar{1}$0]
  ($\blacktriangleleft$), respectively, for
  $\varepsilon_\parallel=\varepsilon_{xy}$ and equivalent directions
  for $\varepsilon_\parallel=\varepsilon_{yz}$.}
\end{figure*}

The calculated strain-dependent MAEs necessary for the determination
of $B_{2}$ are shown in Fig.~\ref{FigNFO_MAE_B2}. The different curves
are adjusted to match at $\varepsilon_\parallel=0$ in order to remove
the corresponding offset which is irrelevant for the present work. The
MAEs are chosen according to Eqs.~\eqref{eq:mae-shear1} and
\eqref{eq:mae-shear2} as energy differences between different in-plane
orientations of the magnetization with respect to the applied shear
strain $\varepsilon_{\parallel}$. According to Eq.~\eqref{EqDeltaEB2}
the slope of the curves given in Fig.~\ref{FigNFO_MAE_B2} is directly
related to the magnetoelastic coupling constant $B_{2}$. From
Fig.~\ref{FigNFO_MAE_B2} it can be seen that the slopes of these
curves are positive, corresponding to positive $B_{2}$. There are
slightly stronger nonlinearities in the curves in each of the panels
compared to Fig.~\ref{FigNFO_MAE_B1}, as well as a stronger influence
of the explicit cation arrangements. The resulting magnetoelastic
coupling constants $B_{2}$ are listed in
Tab.~\ref{TableMagneticPropertiesNFO} and range from 0.9~MPa to
2.5~MPa, leading to magnetostriction constants $\lambda_{111}$ ranging
from $-3.4\times 10^{-6}$ to $-9.7\times 10^{-6}$. These values are
compatible with the lower experimental values, which themselves range
from $-4.0\times 10^{-6}$ to $-23.8\times 10^{-6}$. The last column in
Tab.~\ref{TableMagneticPropertiesNFO} also lists the averaged
$\lambda_{S}$ suitable for polycrystalline materials using
Eq.~\eqref{EqLambdaS}.

Overall, the different cation arrangements and strain planes have only
a rather weak influence on the calculated magnetostriction constants
of NFO, which agree very well with the range of reported experimental
data. We can therefore confirm our earlier
finding,\cite{Fritsch_PRB82_104117} that DFT+$U$ methods are suitable
for a quantitative description of magnetoelastic properties in this
material. Moreover, although the symmetries of the investigated cation
arrangements are not cubic, the magnetostrictive properties of NFO are
very well described within the cubic theory.

\subsubsection{CFO}
\label{Chapter3.2.2}

\begin{figure}
\includegraphics[width=\columnwidth,clip]{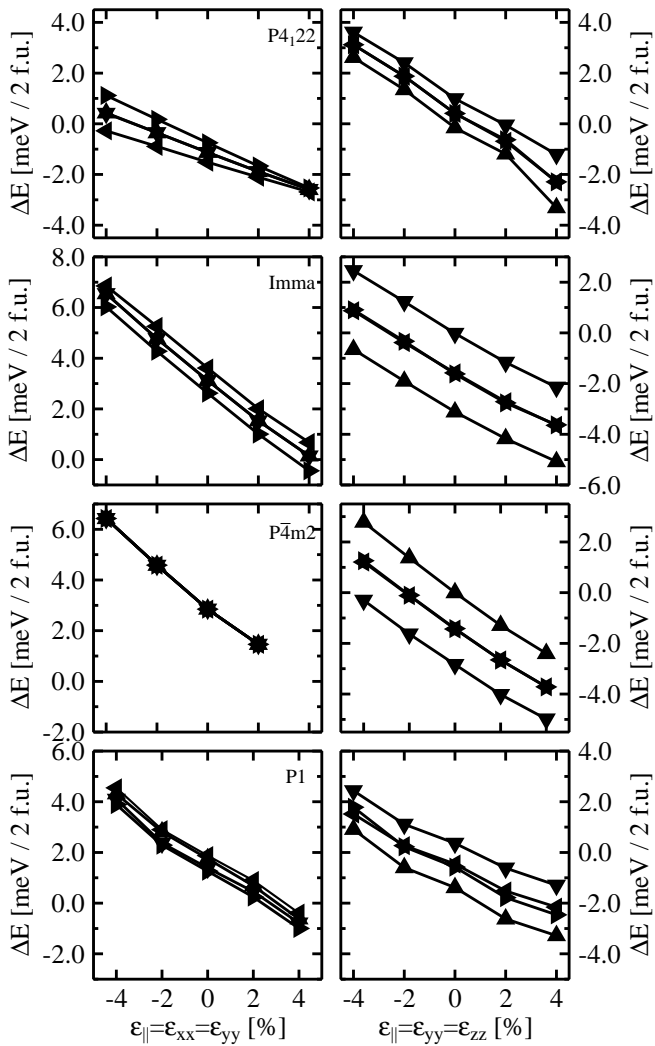}
\caption{\label{FigCFO_MAE_B1} Total energy difference $\Delta E$ per
  two formula units (f.u.) of CFO as function of the epitaxial
  constraint $\varepsilon_{\parallel}$ for different cation
  arrangements. The left (right) panels correspond to the case with
  $\varepsilon_{\perp}=\varepsilon_{zz}$
  ($\varepsilon_{\perp}=\varepsilon_{xx}$). The panels from top to
  bottom refer to symmetries $P4_{1}22$, $Imma$, and $P\bar{4}m2$, and
  $P1$ (low-energy solution for cation inversion
  $\lambda=0.75$\cite{Fritsch_APL99_081916}). In case of
  $\varepsilon_{\perp}=\varepsilon_{zz}$
  ($\varepsilon_{\perp}=\varepsilon_{xx}$) the depicted energy
  difference $\Delta E$ is taken with respect to the [001] ([100])
  direction, with the symbols denoting $\blacktriangle$ [100] ([010]),
  $\blacktriangledown$ [010] ([001]), $\blacktriangleleft$
  [1$\bar{1}$0] ([01$\bar{1}$]), and $\blacktriangleright$ [110]
  ([011]), respectively.}
\end{figure}

Now we turn to our results for CFO. The calculated MAEs for the
determination of the magnetoelastic coupling constant $B_{1}$ are
depicted in Fig.~\ref{FigCFO_MAE_B1}. At first sight, one notices
again that all slopes are negative, leading to a positive
magnetoelastic coupling constant $B_{1}$. However, in contrast to NFO,
the values are now much larger and also depend more strongly on the
specific cation arrangement and orientation of the strain plane.  In
all cases except for the case of $P\bar{4}m2$ with
$\varepsilon_\parallel=\varepsilon_{xx}$, we again obtain an offset
between the different curves, which is due to the lower symmetry of
the specific cation distribution. The differences in slopes observable
between the various curves in the left panel of $P4_{1}22$ symmetry
are due to the fact that in this case the system adopts an
orbitally-ordered ground state with symmetry lower than that of the
underlying crystal structure. Strongest deviations from linearity are
observed in the low-energy solution with symmetry $P1$ belonging to
incomplete inversion $\lambda=0.75$.

\begin{table}
\caption{\label{TableMagneticPropertiesCFO} Magnetoelastic coupling
  constants ($B_1$, $B_2$) and magnetostriction constants
  ($\lambda_{100}$, $\lambda_{111}$, $\lambda_{S}$) for CFO using
  different cation arrangements and strain planes according to
  $\varepsilon_{\perp}=\varepsilon_{zz}=z$
  ($\varepsilon_{\perp}=\varepsilon_{xx}=x$) in comparison with
  available experimental data. The average magnetostriction constant
  $\lambda_{S}$ has been obtained using Eq.~\eqref{EqLambdaS}.}
\begin{ruledtabular}
\begin{tabular}{lcccccccc}
& \multirow{2}{*}{$\varepsilon_{\perp}$} & B$_{1}$ & $\lambda_{100}$ & B$_{2}$ & $\lambda_{111}$ & $\lambda_{S}$ \\
& & (MPa) & ($\times 10^{-6}$) & (MPa) & \multicolumn{2}{c}{($\times 10^{-6}$)} \\
\hline
\multirow{2}{*}{$P4_{1}22$} & z & 18.9 & $-$120.1 & $-$8.4 & 32.9 & $-$28.3 \\
& x & 32.8 & $-$215.0 & $-$ & $-$ & $-$ \\ 
\multirow{2}{*}{$Imma$} & z & 39.7 & $-$251.7 & $-$11.6 & 40.9 & $-$76.1 \\
& x & 29.2 & $-$189.7 & $-$12.2 & 48.8 & $-$ \\
\multirow{2}{*}{$P\bar{4}m2$} & z & 42.0 & $-$272.7 & $-$14.6 & 52.7 & $-$77.5 \\
& x & 30.9 & $-$199.4 & $-$ & $-$ & $-$ \\
\multirow{2}{*}{$P1$} & z & 29.1 & $-$196.3 & $-$7.6 & 28.8 & $-$61.2 \\
& x & 24.2 & $-$160.9 & $-$ & $-$ & $-$ \\
\hline
\multirow{3}{*}{Exp.} & Ref.~\onlinecite{Chen_IEEETransMagn35_3652}\footnotemark[1] & & $-$225.0 & \\
         & Ref.~\onlinecite{Bozorth_PhysRev99_1788}\footnotemark[2] & & $-$250.0 & \\
         & Ref.~\onlinecite{Bozorth_PhysRev99_1788}\footnotemark[3] & & $-$590.0 & & 120.0 & $-$164.0 \\
\end{tabular}
\end{ruledtabular}
\footnotetext[1]{Polycrystalline CoFe$_2$O$_4$.}
\footnotetext[2]{Single crystals with Co$_{1.1}$Fe$_{1.9}$O$_4$ composition.}
\footnotetext[3]{Single crystals with Co$_{0.8}$Fe$_{2.2}$O$_4$ composition.}
\end{table}

The determined magnetoelastic coupling constants $B_{1}$ are given in
Tab.~\ref{TableMagneticPropertiesCFO}, ranging from 18.9~MPa to
42.0~MPa. The largest influence of the strain plane orientation is
observed for $P4_{1}22$ symmetry. Overall, the specific cation
arrangement has a much larger influence on the obtained magnetoelastic
coupling constants in CFO compared to NFO. However, we note that even
though there are pronounced differences between the two different
strain orientations (left and right panels in
Fig.~\ref{FigCFO_MAE_B1}) for the same cation arrangements, the strain
dependence of the various calculated energy differences for the same
strain orientation (different curves within each panel) are very
similar in each case. From expression \eqref{EqDeltaETetyz} for
tetragonal symmetry, we can therefore empirically observe that the
following approximate relationship holds between the various
magnetoelastic coefficients:
\begin{equation}
\label{eq:approx-cubic}
\frac{1}{2} b_{3}(\nu_{2D}+1) \approx b_{21}+b_{22}-\nu_{2D}b_{21} \,
.
\end{equation}
However, since the slopes in the left and right panels of
Fig.~\ref{FigCFO_MAE_B1} differ, the stronger condition
$b_3=b_{22}=-2b_{21}$, which would be valid within cubic symmetry, is
not fulfilled in CFO. The deviation from cubic symmetry caused by the
specific cation arrangements, is therefore more strongly manifested in
the magnetoelastic response of CFO compared to NFO. Nevertheless, the
approximate relation Eq.~\eqref{eq:approx-cubic} indicates that some
residue of the approximate structural cubic symmetry is still present
also in the case of CFO.

The magnetostriction constants of CFO can now be obtained via
Eq.~\eqref{EqLambda100Cub} and using the elastic constants in
Table~\ref{TableElasticProperties}. The resulting values are listed in
Table~\ref{TableMagneticPropertiesCFO} and range from $-120.1\times
10^{-6}$ to $-272.7\times 10^{-6}$. This agrees well with the lower
range of available experimental data, which itself varies between
$-225\times 10^{-6}$ and $-590\times 10^{-6}$.

\begin{figure*}
\includegraphics[width=\textwidth,clip]{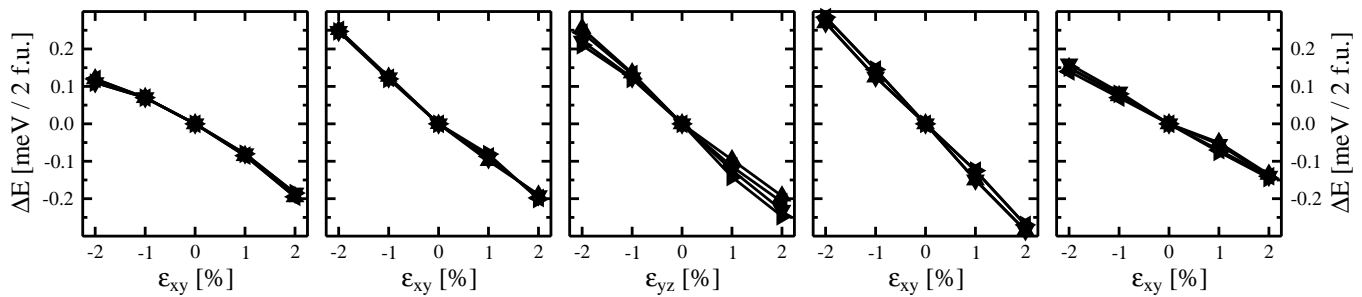}
\caption{\label{FigCFO_MAE_B2} Total energy difference $\Delta E$ per
  two formula units (f.u.) of CFO as function of shear strain for
  different cation arrangements. The panels from left to right refer
  to symmetries $P4_{1}22$
  ($\varepsilon_{\parallel}=\varepsilon_{xy}$), $Imma$
  ($\varepsilon_{\parallel}=\varepsilon_{xy}$), $Imma$
  ($\varepsilon_{\parallel}=\varepsilon_{yz}$), $P\bar{4}m2$
  ($\varepsilon_{\parallel}=\varepsilon_{xy}$), and $P1$
  ($\varepsilon_{\parallel}=\varepsilon_{xy}$, low-energy solution for
  cation inversion $\lambda=0.75$\cite{Fritsch_APL99_081916}),
  respectively. The depicted energy differences $\Delta E$ correspond
  to [110]-[100] ($\blacktriangle$), [110]-[010]
  ($\blacktriangledown$), [100]-[1$\bar{1}$0] ($\blacktriangleright$),
  and [010]-[1$\bar{1}$0] ($\blacktriangleleft$), respectively, for
  $\varepsilon_{\parallel}=\varepsilon_{xy}$ and equivalent directions
  for $\varepsilon_{\parallel}=\varepsilon_{yz}$.}
\end{figure*}

The strain-dependent MAEs necessary for the determination of $B_{2}$
are shown in Fig.~\ref{FigCFO_MAE_B2}, analogous to the NFO case. Most
strikingly, and in contrast to NFO, the corresponding slope is
negative, thus leading to a negative $B_{2}$ in CFO. The spread in
slopes in each of the panels is comparable to NFO. While we obtain
quite similar values for $Imma$ and $P\bar{4}m2$ symmetry (middle
three panels), and also for $P4_{1}22$ and $P1$, the latter two
symmetries lead to somewhat smaller values for $B_2$ than the former.

Overall, $B_{2}$ ranges from $-$8.4~MPa to $-$14.6~MPa for the
symmetries corresponding to complete cation inversion, and $-$7.6~MPa
for the case with $\lambda=0.75$ ($P1$).  The resulting
magnetostriction constants $\lambda_{111}$ of CFO range from
$28.8\times 10^{-6}$ to $52.7\times 10^{-6}$, respectively. These
values are lower than the (to the best of our knowledge only
available) value of $120 \times 10^{-6}$ reported experimentally.

In view of the relatively strong dependence on the specific cation
arrangement, no particular trend is apparent on how the
magnetostriction constants change with reduced cation inversion ($P1$
structure compared to the other cases with full inversion). Taking a
closer look at the individual magnetostriction constants
$\lambda_{100}$ for all investigated cation arrangements and strain
planes, one can notice that the largest magnetostriction occurs for
cases where the cation species are arranged in alternating planes
parallel to the applied strain plane, e.g.,
$\varepsilon_{\perp}=\varepsilon_{zz}=z$ for $Imma$ and $P\bar{4}m2$
symmetry (see Fig.~\ref{FigSpinelCationDistribution}).  Furthermore,
if one compares the two different strain orientations for $P4_{1}22$
symmetry, the magnetostriction is larger for
$\varepsilon_{\perp}=\varepsilon_{xx}=x$, where the strain plane
contains chains of $B$ site cations with two equal cations next to
each other in each chain. The magnetostriction value for the strain
plane containing alternating cation chains within $P4_122$ symmetry is
the smallest observed here. However, at present it is unclear whether
these correlations between cation arrangement and $\lambda_{100}$ are
mostly coincidental, or whether they indeed indicate a deeper
relationship between these two properties. In any case our results
give clear evidence that a fully quantitative model of anisotropy and
magnetostriction in CFO needs to include crystal- or ligand-field
effects that go beyond the immediate nearest neighbor shell of the
Co$^{2+}$ cation.

The effect of different distributions of Co$^{2+}$ and Fe$^{3+}$
cations on the $B$ sites surrounding a specific Co $B$ site has been
taken into account in the theory of magnetic anisotropy for CFO by
Tachiki,\cite{Tachiki:1960} and is also discussed by
Slonczewski.\cite{Slonczewski:1961} It was shown that the
corresponding crystal-field component can have a strong effect on the
resulting cubic magnetic anisotropy constants. The noticeable
dependence of our calculated magnetoelastic coupling constants on the
specific cation arrangement in CFO indicates that this crystal-field
component is indeed quite strong and needs to be taken into account
within a quantitative theory of anisotropy and magnetostriction in
spinel ferrites.

\section{Summary and conclusions}
\label{Chapter4}

In summary, we have presented a detailed first-principles study of
elastic and magnetoelastic properties of the inverse spinel ferrites
NFO and CFO. We have calculated all cubic elastic and magnetoelastic
constants from a variety of distorted crystal structures. Thereby, we
have considered different possible cation arrangements to represent
the inverse spinel structure, and in the case of CFO we also
considered a cation distribution corresponding to incomplete inversion
with $\lambda=0.75$. The magnetoelastic coefficients are obtained from
the strain dependence of the MAEs for two different deformations of
the crystal structure.

Even though the symmetry of the considered cation arrangements is
lower than cubic, our results show that the elastic response of both
NFO and CFO can to a good approximation be described using cubic
elastic constants. Since the elastic constants are mainly determined
by the strength of the chemical bonding, this indicates that Co, Ni,
and Fe all form bonds of similar strength with the surrounding atoms.

Similarly, the magnetoelastic response of NFO can also to a good
approximation be described using the cubic expression for the
magnetoelastic energy density (Eq.~\eqref{EqFMeCub}). This is
indicated by the relatively small quantitative differences in the
calculated magnetoelastic coefficients for the various cation
arrangements. On the other hand, the magnetoelastic coefficients of
CFO show a stronger dependence on the specific cation arrangement and
the orientation of the applied strain, so that the cubic
approximations is less justified in that case. In addition, the
overall magnetoelastic response is much stronger in CFO than in NFO.

Both of these observations can be understood from the $d^7$ electron
configuration of the Co$^{2+}$ cation, which leads to stronger
spin-orbit effects compared with the $d^8$ configuration of
Ni$^{2+}$. In the latter, the orbital magnetic moment is strongly
quenched by the dominant octahedral component of the crystal-field,
and the system is less sensitive to additional crystal field
components of lower symmetry. In contrast, the orbital moment is not
fully quenched by the octahedral crystal field for the $d^7$
configuration of Co$^{2+}$, and additional splittings, which are
created by the different arrangements of the surrounding $B$ site
cations, can have much stronger effects on the electronic ground state
within the partially filled minority-spin $t_{2g}$ orbital manifold.

Both sign and magnitude of the calculated magnetostriction constants
agree well with available experimental data. Even for CFO, where the
calculated magnetostriction depends more strongly on the specific
cation distribution than for NFO, the resulting uncertainty is within
the spread of available experimental data.

Further experimental data for single crystals is therefore required
for a more accurate comparison. We note that a number of obstacles can
in principle affect an accurate comparison between theory and
experiment. Apart from potential influences of varying sample
stoichiometry, degree of inversion, and measuring temperature, the
preparation of an ideal demagnetized state with an essentially random
orientation of magnetic domains is relatively hard to achieve. For
example, a state with 50~\% of domains oriented parallel and 50~\% of
domains oriented antiparallel with respect to a certain axis would
have zero magnetization but the magnetostrictive strain would already
be saturated along that direction. Furthermore, for systems with very
strong magnetic anisotropy, such as e.g. CFO, it can be very difficult
to achieve full saturation along the hard
direction.\cite{Kriegisch_et_al:2012} Other sources of disagreement
between theory and experiment could be due to the neglect of higher
order terms in the energy expression
\eqref{EqFMeCub},\cite{Lee_RepProgPhys18_184} or most likely due to
deficiencies in the exchange correlation potential used in the DFT
calculations. However, based on the currently available experimental
data it can be concluded that the GGA+$U$ method used in the present
work is sufficiently accurate for further investigation on the effects
of cation distribution, degree of inversion, and stoichiometry on the
magnetostrictive properties of spinel ferrites.

Our work thus provides a sound basis for future investigations of
magnetostriction and anisotropy in spinel ferrites as well as for
future first principles studies of magnetoelectric coupling in
artificial multiferroic heterostructures containing either CFO or NFO
in combination with ferroelectric and/or piezoelectric materials.

\begin{acknowledgments}
This work was done mostly within the School of Physics at Trinity
College Dublin, supported by Science Foundation Ireland under
Ref.~SFI-07/YI2/I1051 and made use of computational facilities
provided by the Trinity Centre for High Performance Computing (TCHPC)
and the Irish Centre for High-End Computing (ICHEC).
\end{acknowledgments}


\begin{thebibliography}{39}
\bibitem[{foo({\natexlab{a}})}]{footnote1}
\bibinfo{note}{We note that these expressions are in principle only valid for
  the cubic pointgroups $m3m$, $\bar{4}3m$, and $432$.}
\bibitem[{foo({\natexlab{b}})}]{footnote2}
\bibinfo{note}{Note that we are not considering any strain indendent
  contributions to the magnetic anisotropy energy, which would lead to (strain
  independent) differences between inequivalent in-plane directions.}
\end{thebibliography}
\end{document}